\documentclass[%
reprint,
onecolumn,
amsmath,amssymb,
aps,
prfluids,
longbibliography,
superscriptaddress,
]{revtex4-2}
\usepackage{booktabs}
\usepackage{graphicx,subfigure}
\usepackage{dcolumn}
\usepackage{float}
\usepackage{bm}
\usepackage[utf8]{inputenc}

\usepackage{xcolor}\usepackage{xcolor}

\begin{document}

\preprint{APS/123-QED}

\title{Hydrodynamics of molecular rotors in lipid membranes}
\author{Vinny Chandran Suja}
\altaffiliation{These authors contributed equally to this work.}\affiliation{%
School of Engineering and Applied Sciences, Harvard University, Cambridge, MA 02138}
\author{Naomi Oppenheimer}
\altaffiliation{These authors contributed equally to this work.}\affiliation{%
 Faculty of Exact Sciences, Tel Aviv University, Tel Aviv-Yafo, 6997801, Israel
}
\author{Howard A. Stone}
 \email{hastone@princeton.edu}
\affiliation{%
Department of Mechanical and Aerospace Engineering,
Princeton University, Princeton, New Jersey 08544, USA
}

\date{\today}

\begin{abstract}

Molecular rotors form twisted conformations upon photoexcitation, with their fluorescence relaxation time serving as a measure of viscosity. They have been used to assess membrane viscosities but yield higher values compared to other methods. Here, we show that the rotor's relaxation time is influenced by a combination of membrane viscosity and interleaflet friction. We present a theory for the relaxation time and obtain a correction factor that accounts for the discrepancy. If the membrane's viscosity is known, molecular rotors may enable the extraction of the elusive interleaflet friction.

\end{abstract}

\maketitle
\section{Introduction}

Biological membranes encase cells and subcellular structures in living organisms, serving as barriers that regulate macromolecule transport, cell adhesion, mechanotransduction, and communication \cite{petty2013molecular}. Biological membranes also facilitate enzymatic and metabolic activities vital for cellular processes. These critical functions of membranes are 
dependent on their biophysical characteristics. However, despite extensive research, much remains to be discovered about important biophysical properties such as viscosity \cite{singer1972fluid, henle2009effective, oppenheimer2009correlated, fitzgerald2023surface, venable2017lipid, shi2022hydrodynamics, huang2024interfacial} and interleaflet friction \cite{evans1994hidden}. 

Conventional rheometry techniques are not convenient to measure the viscosity of a membrane bilayer, especially at {\it in vivo} length scales. Instead, more intricate, microscopic methods are used. One such method is Fluorescence Recovery After Photobleaching (FRAP) \cite{edidin1976measurement} --- lipids are marked by a fluorescent dye, a small area is photobleached, and the recovery of the fluorescence in the photobleached section is followed in time. Analyzing the fluorescence recovery kinetics, yields the Brownian diffusion coefficient ($D_T$) of the lipids, which is related to lipid translational resistance ($\Lambda$) through the Einstein relationship $D_T = k_B T/ \Lambda_T$.  For the simplest configurations, the membrane viscosity is inferred from resistance using the Saffman-Delbr\"{u}ck approximation \cite{saffman1975brownian}, $\Lambda_T = 4\pi\eta_m/ \left [\ln\left (\frac{\eta_m}{\eta R}\right ) -  \gamma\right ]$, where $\eta_m$ is the two-dimensional (2D) surface viscosity of the membrane, which can be related to a thin film viscosity $\eta_m^*$  as $\eta_m = \eta_m^* h$ \cite{evans1988translational}; $h$ is the thickness of the membrane, $\eta$ is the three-dimensional (3D) viscosity of the surrounding fluid, $R$ is the radius of the diffusing particle, and $\gamma \simeq 0.5772$ is Euler's constant. This formula for $D_T$ assumes  $\frac{\eta_m}{\eta R} \gg 1$ so that $D_T > 0$. 

A relatively new method for measuring membrane viscosity involves the use of so-called molecular rotors~\cite{haidekker2007molecular, nipper2008characterization, levitt2009membrane, wu2013molecular, lopez2014molecular, dent2015imaging}. When these molecules are photoexcited, they form twisted intramolecular charge transfer states. Following excitation, the rotors relax via a combination
of two competing mechanisms: 1) fluorescence and 2) non-radiative untwisting. A more viscous fluid retards the rate of relaxation via untwisting, which  leads to relaxation mainly by fluorescence~\cite{forster1971viskositatsabhangigkeit, haidekker2010environment}. The fluorescence lifetime and intensity is therefore an indication of the viscosity of the medium. In particular, for an intermediate range of viscosities (usually between $0.01$ -- $1$ Pa$\cdot$s), the fluorescence lifetime in the bulk, $\tau_{f,3D}$, and bulk viscosity, $\eta$, are related by a power-law relationship, 
\begin{equation}
 \tau_{f,3D} = \frac{z\eta^\alpha}{k_r} 
 \label{BulkFluorescenceLifetime}
\end{equation}where $k_r$ is the radiative decay rate, and $z$ and $\alpha$ are constants \cite{haidekker2010environment, kuimova2012mapping}. The unknown constants are usually obtained by calibrating rotor lifetimes using a series of liquid mixtures with known bulk viscosity \cite{dent2015imaging, singh2023molecular}. Following calibration, Eq.\ (\ref{BulkFluorescenceLifetime}) is used to recover thin film membrane viscosity $\eta_m^*$ from fluorescence lifetimes measured in lipid membranes.

Membrane viscosities obtained via fluorescence lifetimes of molecular rotors are, however, consistently larger than those independently obtained through diffusion measurements utilizing the Saffman-Delbr\"{u}ck formula \cite{dent2015imaging, adrien2022best, wu2013molecular, nipper2008characterization}. Although several potential reasons have been proposed to explain this discrepancy, no quantitative resolutions are available. While molecular rotors are not expected to yield viscosity measurements comparable to those obtained from conventional rheometers, in 3D, molecular rotor relaxation kinetics are predicted well by continuum hydrodynamics theories \cite{forster1971viskositatsabhangigkeit}, and they yield accurate viscosity measurements. By extension, we hypothesize that molecular rotors will yield accurate measurements of membrane viscosities in 2D, provided the hydrodynamics of molecular rotors within membranes is accurately described. In alignment with this hypothesis, we present a theoretical hydrodynamic basis  to address the discrepancy in membrane viscosities measured by molecular rotors by accounting for differences between rotor relaxation kinetics during calibration and during measurements within a membrane. Notably, we show that when embedded in membranes, molecular rotors measure a combination of the two-dimensional (2D) membrane viscosity and another difficult-to-measure quantity, interleaflet friction \cite{anthony2022systematic,zgorski2019surface, camley2013diffusion}.

\begin{figure}[t!]
    \centering\includegraphics[width=0.6\linewidth]{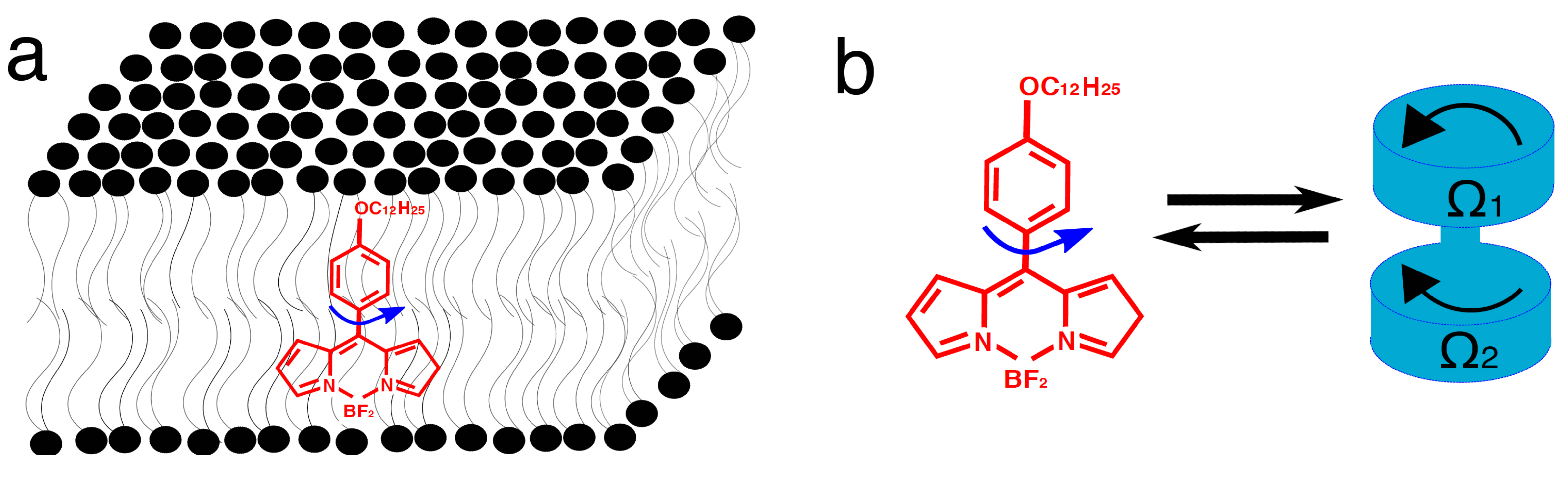}
    \caption{A membrane-bound molecular rotor. (a) A representation of a molecular rotor, in this case a BODIPY rotor, in a biological membrane. (b) Our simplified model of a molecular rotor is composed of two counter-rotating disks, one with angular velocity $\Omega_1$ in the upper leaflet, and the second with angular velocity $\Omega_2$ in the bottom leaflet.}
    \label{fig:fig1}
\end{figure}

Current understanding of the position of the molecular rotor within the bilayer (e.g.,  \cite{dent2015imaging}; see also Fig. \ref{fig:fig1}) has it roughly spanning the mid-plane, with the bulk of the counter-rotating moieties of the rotor differentially localized to each of the bilayer leaflets (see SI for additional details). This implies that when the molecule twists it induces shear between the two lipid layers in addition to rotational flows in each leaflet. With this picture in mind, we provide a theoretical prediction for the relaxation time of an initially twisted molecule as a function of both the membrane viscosity and interleaflet friction. We compare our theory to results given in the literature and show that the theoretical predictions can explain the discrepancy in viscosity measurements.  As an additional outcome, it may be possible to extract interleaflet friction from measurements with molecular rotors if the membrane's viscosity is known by other means, such as from FRAP or Fluorescence Correlation Spectroscopy (FCS) measurements. To proceed, 
we first present the problem of two counter-rotating disks in a membrane and find the typical relaxation time of an initially twisted molecular rotor. 
Then, we discuss the results and use them to reinterpret existing experiments in the literature.

\section{Theory}\label{sec:Theory}
In order to determine the relaxation time of an excited molecular rotor in a membrane we will make a considerable simplification: we assume the molecular rotor is made of two counter-rotating disks, one in each leaflet (see Fig. \ref{fig:fig1}). As outlined below
the effective velocity of two counter-rotating disks can be interpreted as a combination of two axillary problems  discussed below: 1) the velocity due to a disk rotating in a viscous 2D flow configuration following the original model of Saffman and Delbr\"uck \cite{saffman1975brownian}, and 2) the flow due to a disk rotating in a 2D ``Brinkman fluid"~\cite{evans1988translational}, where there is additional friction on the leaflet, as in the case of a supported bilayer \cite{sackmann1996supported, stone1998hydrodynamics,oppenheimer2010correlated}. The application of a continuum hydrodynamic theory to predict rotor dynamics is consistent with well-established theories including the Saffman-Delbr\"{u}ck model \cite{saffman1975brownian}, F\"orster-Hoffmann model \cite{forster1971viskositatsabhangigkeit}, and the Stokes-Einstein-Debye model \cite{sluch2002friction}, all of which invoke continuum hydrodynamic arguments down towards molecular scale to obtain hydrodynamic resistances, as we aim to do below (see SI for additional details).

\begin{figure}[ht!]
    \centering
    \includegraphics[width=0.6\linewidth]{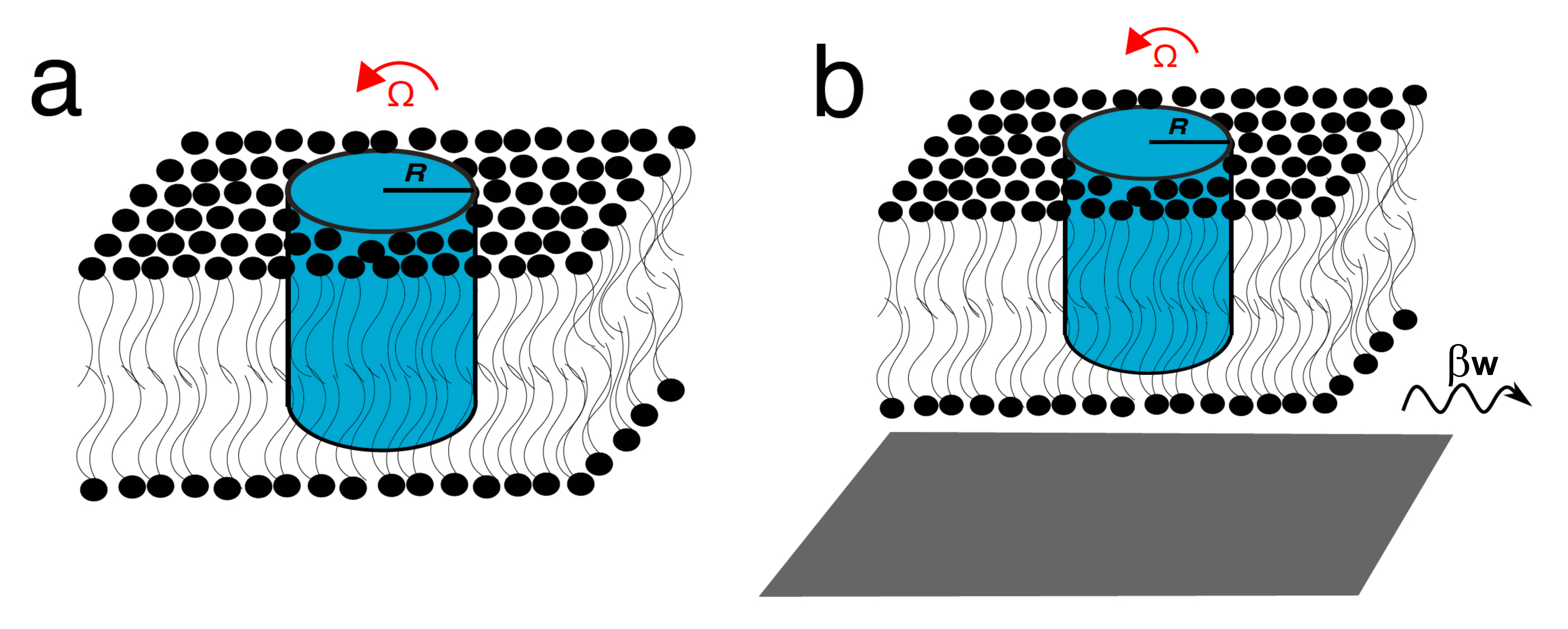}
    \caption{Hydrodynamic characterization. (a) A rotating disk of radius $R$ with angular velocity $\Omega$ in a biological membrane surrounded on both sides by an infinite outer fluid. The leaflets are assumed to have the same properties. (b) A rotating disk of radius $R$ and angular velocity $\Omega$ in a membrane that is in the vicinity of a rigid wall. The fluid dissipates momentum to the wall with a friction coefficient $b_w$.}
    \label{fig:fig2}
\end{figure}

\subsubsection{Rotating disk in a two-dimensional viscous fluid}. \label{sec:RotatingDisk2D} 
Consider a disk of radius $R$ rotating with an angular velocity $\Omega$ in a 2D viscous fluid of viscosity $\eta_m$. The velocity field $\mathbf{v}$ in the membrane is governed by the Stokes equations, $\eta_m \nabla^2 \mathbf{v} = \nabla p$, where $p$ is the pressure field. Here we have neglected the influence of the surrounding fluid on the membrane as $h \eta/\eta_m \ll 1$.For characteristic values of variables reported in Fig. \ref{fig:ViscosityDiffusionLifetime}, $h \eta/\eta_m$ is less than $3\times10^{-2}$. Furthermore, in 2D, the velocity field generated by rotation falls off faster (varies as $1/r$, see Eq.\ref{vStokes}) than that generated by translation (varies as $\log r$) \cite{saffman1975brownian}. As a consequence, viscous dissipation is localized to the membrane and the influence of bulk fluid viscosity on rotational drag can be safely neglected. From symmetry, we can assume that the flow field is in the $\mathbf{e}_\theta$ direction and is a function of $r$ alone, such that $\mathbf{v} = v(r) \mathbf{e}_\theta$. For a solution of this form the incompressibility requirement is implicitly satisfied and there are no gradients of pressure. The equation of motion is thus the Laplace equation:
\begin{equation}
    \nabla^2 \mathbf{v} = 
    {\bf 0},
\end{equation}which is to be solved 
with boundary conditions,
\begin{equation} \label{eq:BoundaryConditionCylinder2D}
    \mathbf{v}(r=R) = \mathbf{\Omega} \times \mathbf{R} = \Omega R \mathbf{e}_\theta  \quad \text{and} \quad \mathbf{v}(r \rightarrow\infty) = {\bf 0}.
\end{equation}The solution is
\begin{equation}
\label{vStokes}
    \mathbf{v}(r) = \frac{\Omega R^2}{r} \mathbf{e}_\theta.
\end{equation}

A straightforward calculation from the surface shear stress, $\sigma_{r\theta}$, integrated over the disk surface, yields the well-known  rotational resistance of a cylinder in a 2D membrane \cite{saffman1975brownian},
\begin{equation}
\label{rStokes}
    \Lambda_{R,\text{I}} = 4\pi \eta_m R^2.
\end{equation}

\subsubsection{Rotating disk in a two-dimensional Brinkman fluid}. \label{sec:RotatingDiskBrinkman2D}
Next consider a disk that is rotating with angular velocity $\Omega$ in a 2D Brinkman fluid, such as for the case of a membrane close to a rigid wall. This flow model introduces an effective force on the membrane due to viscous stresses from the surrounding fluid. We follow Evans and Sackmann \cite{evans1988translational} in writing the equation of motion,
\begin{equation}
    \eta_m \nabla^2 \mathbf{v} -  \beta_w \mathbf{v} = {\bf 0} \quad \text{with} \quad \mathbf{v} = v(r) \mathbf{e}_\theta,
\end{equation}
where $\beta_w$ is the friction coefficient with the wall. The boundary conditions are similar to Eq. (\ref{eq:BoundaryConditionCylinder2D}). The solution is
\begin{equation}
\label{vBrinkman}
    \mathbf{v}(r) = \frac{\Omega R K_1 (\gamma r)}{K_1 (\gamma R)} \mathbf{e}_\theta,
\end{equation}where $K_i(\cdot )$ is the modified Bessel function of the second kind of order $i$ and $\gamma^2 = \beta_w/\eta_m$. The rotational resistance follows as \cite{evans1988translational}
\begin{equation}
\label{rBrinkman}
    \Lambda_{R,\text{II}} = \left (4\pi \eta_m R^2\right ) \left(1 + \frac{\gamma R K_0(\gamma R)}{2K_1(\gamma R)} \right) = \frac{2 \pi \eta_m R^3 \gamma K_2(\gamma R)}{K_1(\gamma R)} .
\end{equation}
 Momentum is conserved up to distances $1/\gamma$ and at larger distances is lost due to friction, such that,  in the limit $r\ll\gamma^{-1}$, Eq.~(\ref{vBrinkman}) converges to Eq.~(\ref{vStokes}) and Eq.~(\ref{rBrinkman}) converges to Eq.~(\ref{rStokes}).

\begin{figure}[th!]
    \centering
    \includegraphics[width=0.5\linewidth]{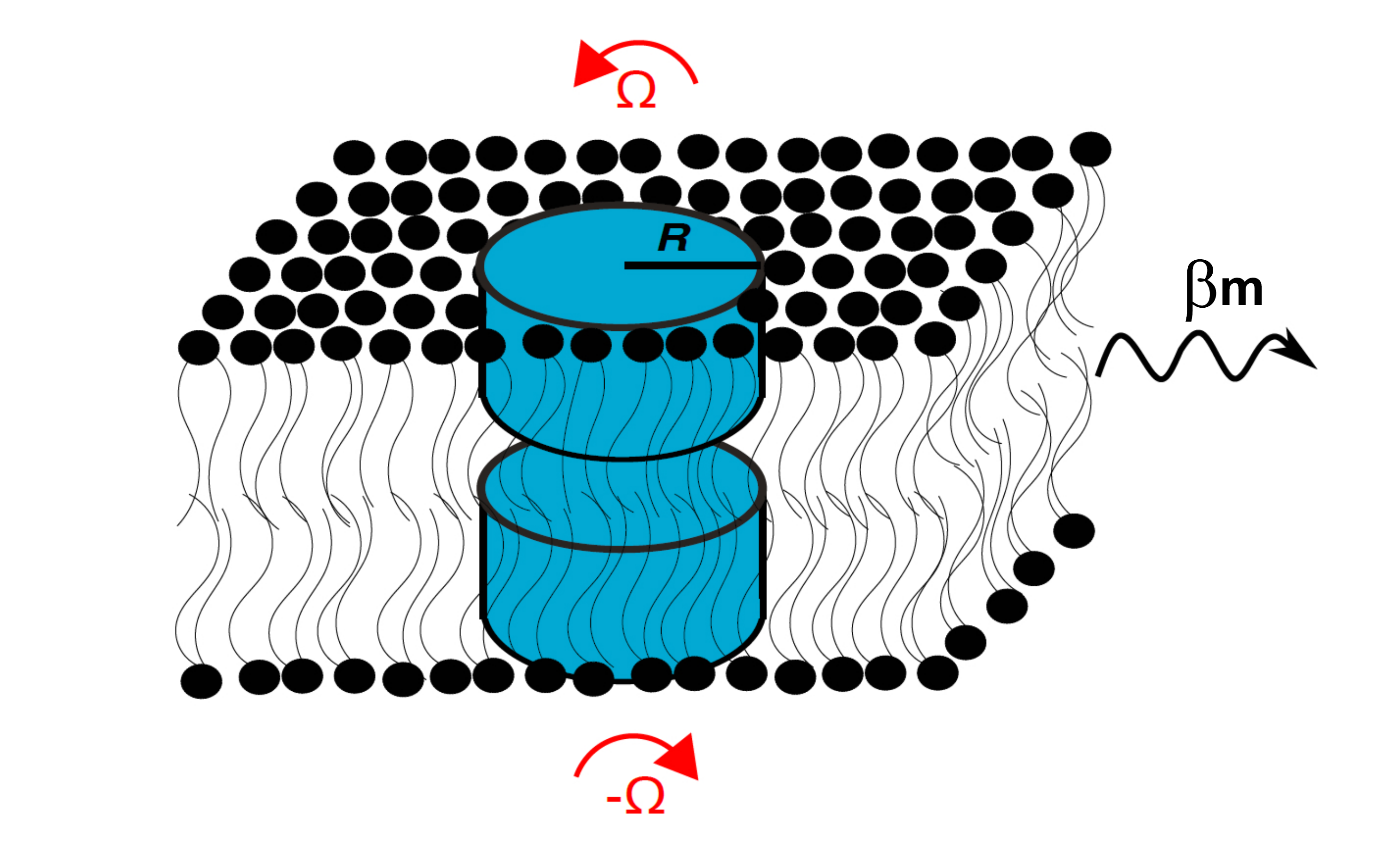}
    \caption{Two counter-rotating disks of radius $R$, respectively with angular velocities $\Omega_1$ and 
$\Omega_2$, in a membrane surrounded on both sides by an infinite outer fluid. We denote the friction between the layers of the membrane as $\beta_m$.}
    \label{fig:fig3}
\end{figure}
\subsubsection{Two counter-rotating disks in a membrane} \label{subsec:CounterRotatingDisks}
Now consider a molecular rotor composed of two connected disks each of radius $R$ immersed between two membrane leaflets. The top disk rotates with angular velocity $\Omega_1$ and the bottom disk rotates in the opposite direction  with angular velocity $\Omega_2$ (see Fig. \ref{fig:fig3}). Suppose that the interleaflet friction is $\beta_m$ and as before assume that $\beta_m$ and $\eta_m$ are large enough to neglect the stress coming from the outer fluid. The equations of motion for such a case can be written as,
\begin{align} \label{eq:CounterRotatingDisks}
    \eta_m \nabla^2 \mathbf{v_1} - \beta_m (\mathbf{v_1 - v_2}) = {\bf 0}, \\
    \eta_m \nabla^2 \mathbf{v_2} - \beta_m (\mathbf{v_2 - v_1}) = {\bf 0},\nonumber
\end{align}where $v_1\;(v_2)$ is the velocity in the upper (lower) leaflet. The boundary conditions are 
\begin{align}\label{eq:CounterRotatingDisksBC}
\mathbf{v_1}(R) = \Omega_1 R \mathbf{e}_\theta \quad \text{and} \quad   \mathbf{v_1}(r \to \infty ) = {\bf 0},\\
\mathbf{v_2}(R) = \Omega_2 R \mathbf{e}_\theta \quad \text{and} \quad   \mathbf{v_2}(r \to \infty ) = {\bf 0} .\nonumber
\end{align}
We can add and subtract Eqs.~(\ref{eq:CounterRotatingDisks}) and the boundary conditions in order to obtain expressions for the joint velocity $\mathbf{U} = \mathbf{v_1} + \mathbf{v_2} = U(r) \mathbf{e}_\theta$ and the relative velocity $\mathbf{V} = \mathbf{v_1} - \mathbf{v_2} = V(r) \mathbf{e}_\theta$. The joint velocity is similar to Eq.~(\ref{vStokes}),
\begin{equation}
    \mathbf{U}(r) = \frac{(\Omega_1 + \Omega_2)R^2}{r} \mathbf{e}_\theta .
\end{equation}
The relative velocity satisfies a governing equation similar to the Brinkman case,  Eq.~(\ref{vBrinkman}), 
with the solution,
\begin{equation}
    \mathbf{V}(r) = (\Omega_1 - \Omega_2)R \frac{K_1(\kappa r)}{K_1(\kappa R)} \mathbf{e}_\theta,
\end{equation}
where $\kappa R$ is a non-dimensional radius defined as $\left(\frac{2\beta_m}{\eta_m}\right)^{1/2}R$.

If the molecule has an initial twist, then it will relax back to equilibrium. In particular, where there is no net torque acting on the molecule, conservation of angular
momentum dictates $\Omega_1 = -\Omega_2 = \Omega$, such that $U = 0$ and 
\begin{equation}
    V(r) = \frac{\dot{\theta}R K_1(\kappa r)}{K_1(\kappa R)},
\end{equation}
where we used the fact that the relative angular velocity of the two disks is equal to the time-rate-of-change of the change of twist in the molecule, $2\Omega = \dot{\theta}$. The above equation also reveals that molecular rotors, while undergoing small angular displacements of approximately $1\; nm$, can induce long range velocity disturbances that span several molecular radii across the membrane (see SI). The effective rotational resistance of the molecule can be obtained by computing the ratio of the net hydrodynamic torque to the relative angular velocity $\dot{\theta}$ as,
\begin{equation} \label{rRotor}
\Lambda_R = \left (4\pi \eta_m R^2\right )\left(1 + \frac{\kappa R K_0(\kappa R)}{2K_1(\kappa R)} \right) = \frac{2 \pi \eta_m R^3 \kappa K_2(\kappa R)}{K_1(\kappa R)}.
\end{equation}
In the limit $R\ll\kappa^{-1}$, Eq.~(\ref{rRotor}) converges to Eq.~(\ref{rStokes}).

\begin{figure*}[!t]
    \centering
    \includegraphics[width =
    0.9\linewidth]{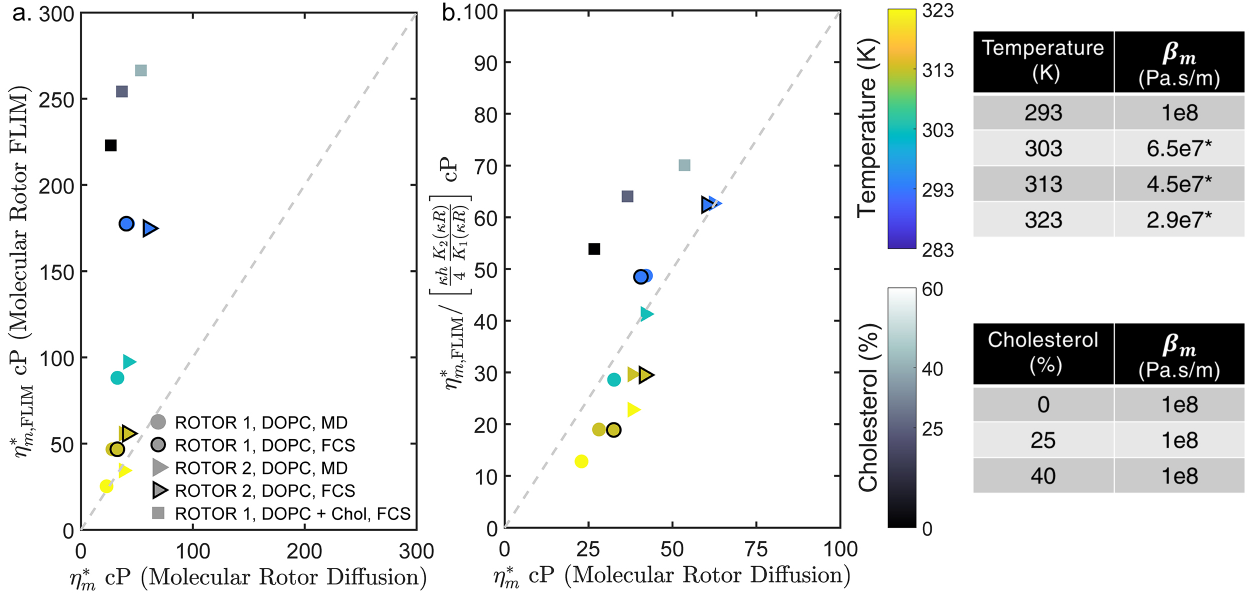}
    \caption{ Comparison with experiments. (a) Membrane viscosity evaluated from molecular translational diffusivity measurements ($\eta_m^*$) is consistently lower than those obtained from fluorescence lifetime measurements ($\eta^*_{m,\rm FLIM}$). Data from DOPC vesicles reported by Dent et al. \cite{dent2015imaging} at different temperatures, and by Wu et al. \cite{wu2013molecular} for different cholesterol concentrations at room temperature. Reported rotor diffusion measurements were obtained via Fluorescence Correlation Spectroscopy (FCS) or Molecular Dynamics (MD). (b) The fold differences between $\eta^*_{m,\rm FLIM}$ and $\eta_m^*$ are in alignment with those expected from Eq. (\ref{eq:ViscosityFoldDifference}). The values of $\beta_m$ used for the calculations are shown on the far right. $\beta_m$ at 293K was obtained from Ref.\ \cite{evans1976membrane}. The same value was used at different cholesterol concentrations as $\beta_m$ is known to have a weak dependence on cholesterol concentration \cite{anthony2022systematic}. As reliable data for $\beta_m$ in DOPC vesicles was not available at other temperatures, $\kappa$ was assumed to remain the same across temperature. A * is used to indicate the $\beta_m$ satisfying this assumption. This choice is reasonable as $\kappa \propto \sqrt{\beta_m/\eta_m}$ should only have a weak temperature dependence as both $\beta_m$ and $\eta_m$ fall with temperature. See SI for additional details. }
    \label{fig:ViscosityDiffusionLifetime}
\end{figure*}

\subsubsection{Fluorescence lifetime of a molecular rotor in a membrane}.
Following F\"{o}rster and Hoffmann \cite{forster1971viskositatsabhangigkeit}, we neglect inertia of the molecule and express the angular relaxation of the rotor through a spring-dashpot-like response, $\Lambda_R \dot{\theta} + k (\theta - \theta_0)  = 0$. Here the spring constant $k$ is governed by molecular-scale interactions that drive the molecule back to its equilibrium angular orientation ($\theta_0$). The solution yields a classical decaying exponential with a relaxation time constant, $t_{2D}^*$, given by
\begin{equation}\label{eq:2DAngularRelaxation}
    t_{2D}^* = \frac{\Lambda_R}{k} =  \frac{2 \pi \eta_m R^3 \kappa K_2(\kappa R)}{k\,K_1(\kappa R)}.
\end{equation}

To link the angular relaxation timescale of the rotor to its fluorescence lifetime in the membrane ($\tau_{f,2D}$), we again follow F\"{o}rster and Hoffmann \cite{forster1971viskositatsabhangigkeit}, and assume that the probability of a molecular rotor occupying its excited state is governed by two competing processes: 1) radiative deactivation with a constant lifetime and 2) conformation dependent non-radiative deactivation. Computing the total quantum yield ($\Phi_f$) as the time integral of molecular excitation probability  yields a relationship linking $\tau_{f,2D}$ and $t_{2D}^*$ (see SI),  

\begin{equation}\label{eq:2DFluorescenceLifetime}
    \Phi_f = \tau_{f,2D} k_r  \propto t_{2D}^{*2/3} \propto \Lambda_R^{2/3} \propto \eta_m^{2/3},
\end{equation}
where $k_r$ is the radiative decay rate introduced in Eq.~(\ref{BulkFluorescenceLifetime}). Note that Eq.~(\ref{eq:2DFluorescenceLifetime}) broadly links fluorescence lifetime with the rotor angular relaxation timescale and is also valid in the bulk \cite{forster1971viskositatsabhangigkeit}.

\section{Discussion}\label{sec:Discussion}
Membrane viscosities obtained via Fluorescence Lifetime Imaging Microscopy (FLIM) measurements are consistently larger than those obtained through diffusivity measurements. This fact can be seen in Fig. \ref{fig:ViscosityDiffusionLifetime}a, where available data in the literature from DOPC vesicles, including results obtained at different temperatures and with varying cholesterol fractions, are plotted to show the systematic overestimation of membrane viscosity by FLIM. The discrepancy can be quantitatively explained solely from differences between rotor hydrodynamics during calibration and during measurements within a membrane. 

Molecular rotors are usually calibrated by correlating fluorescence lifetimes measured in the bulk ($\tau_{f,3D}$) with the bulk viscosity ($\eta$) across a series of liquid mixtures~\cite{dent2015imaging, wu2013molecular, singh2023molecular}. From Eq. (\ref{eq:2DFluorescenceLifetime}), it follows that $\tau_{f,3D} \propto t^{*2/3}_{3D}$, the angular relaxation time constant of the rotor in the bulk. Employing the well known rotational resistance of spheres in 3D, $t^{*}_{3D}$ can be shown to be equal to $8\pi\eta R^3/k$ (see SI). Comparing $t^{*}_{3D}$ with the result from Eq.\ (\ref{eq:2DAngularRelaxation}) that explicitly accounts for the effects of interleaflet friction on rotor hydrodynamics in membranes yields   
\begin{equation}\label{eq:ViscosityFoldDifference}
 \frac{t_{2D}^*}{t_{3D}^*} = \frac{\eta_m \kappa K_2(\kappa R)}{4\eta K_1(\kappa R)} = \left(\frac{\tau_{f,2D}}{\tau_{f, 3D}} \right)^{3/2}.
\end{equation}
In the literature, membrane viscosity is usually reported as a thin film viscosity ($\eta_m^*$) with dimensions of bulk viscosity \cite{dent2015imaging, wu2013molecular, singh2023molecular}. $\eta_m^*$ is related to the 2D membrane viscosity $\eta_m$ as $\eta_m = \eta_m^* h$ \cite{evans1988translational}, where $h$ is the thickness of the membrane. Making this substitution in Eq. (\ref{eq:ViscosityFoldDifference}), we see that simply equating fluorescence lifetimes measured on a membrane to those obtained from calibration experiments in the bulk can overestimate membrane thin film viscosity ($\eta_m^*$) by a factor of $\kappa h K_2(\kappa R)/4 K_1(\kappa R)$. 

Rescaling the FLIM membrane viscosity data in Fig. \ref{fig:ViscosityDiffusionLifetime}a with $\kappa h K_2(\kappa R)/4 K_1(\kappa R)$ leads to a significantly improved agreement between membrane viscosities from the two different measurements (Fig.~\ref{fig:ViscosityDiffusionLifetime}b), where the values of $\beta_m$ (for calculating $\kappa$) and $R$ are obtained from the literature (see SI). Physically, the identified factor accounts for two aspects that were previously overlooked. First, the hydrodynamics of molecular rotor relaxation in lipid membranes is affected by the interleaflet friction ($\beta_m$) in addition to membrane viscosity. This is accounted for by $\kappa$. Second, the 3D hydrodynamics of molecular rotor relaxation in the bulk (experienced during calibration) is different from the 2D hydrodynamics in a thin lipid bilayer (experienced during measurement). The key variables influencing this discrepancy can be isolated by taking the limit $\kappa \rightarrow 0$ in Eq. (\ref{eq:ViscosityFoldDifference}), whereby the factor simplifies to $h/(2R)$.



These results underscore two key takeaways. Calibration curves obtained via 3D viscosity measurements should be corrected by multiplying the lifetime by a factor of   
$\left[\kappa h K_2(\kappa R)/4 K_1(\kappa R)\right]^{2/3}$ for directly obtaining the accurate membrane viscosity.
Second, if membrane viscosity $\eta_m^*$ is independently available, e.g., FCS, FRAP or MD simulations, Eq.\ (\ref{eq:ViscosityFoldDifference}) provides a convenient way to infer interleaflet friction  --- a hard-to-measure quantity, particularly on curved liposomes and {\it in vivo}. In this case, the interleaflet friction can be numerically extracted by solving $\eta^*_{m,\rm FLIM}/\eta_m^* =  \kappa h K_2(\kappa R)/4 K_1(\kappa R)$, where $\eta^*_{m,\rm FLIM}$ is the uncorrected membrane viscosity (obtained conventionally from bulk viscosity calibrated fluorescence lifetimes). These results also suggest that molecular rotors with a larger radius are better suited for measuring interleaflet friction, thus providing guidance on the development of rotors optimized for sensing interleaflet friction (see SI). With further investigation on more molecular rotors and lipid systems, the provided framework can 
expand the use of molecular rotors as valuable molecular rheometry probes. 


\section*{Acknowledgements}
\noindent We note that this work began in 2018 when NO was a postdoc with HAS, but stalled because the researchers were unable to see the connection of the theory to the experimental literature. A conversation between VCS and HAS in Spring 2024 led VCS to reconsider the findings and demonstrate the ability of the theory to rationalize the existing experimental measurements.
\bibliographystyle{vancouver}
\bibliography{References}

\begin{thebibliography}{10}

\bibitem{petty2013molecular}
Petty HR.
\newblock Molecular Biology of Membranes: Structure and Function.
\newblock Springer Science \& Business Media; 2013.

\bibitem{singer1972fluid}
Singer SJ, Nicolson GL.
\newblock The Fluid Mosaic Model of the Structure of Cell Membranes: Cell membranes are viewed as two-dimensional solutions of oriented globular proteins and lipids.
\newblock Science. 1972;175(4023):720-31.

\bibitem{henle2009effective}
Henle ML, Levine AJ.
\newblock Effective viscosity of a dilute suspension of membrane-bound inclusions.
\newblock Physics of Fluids. 2009;21(3).

\bibitem{oppenheimer2009correlated}
Oppenheimer N, Diamant H.
\newblock Correlated diffusion of membrane proteins and their effect on membrane viscosity.
\newblock Biophysical Journal. 2009;96(8):3041-9.

\bibitem{fitzgerald2023surface}
Fitzgerald JE, Venable RM, Pastor RW, Lyman ER.
\newblock Surface viscosities of lipid bilayers determined from equilibrium molecular dynamics simulations.
\newblock Biophysical Journal. 2023;122(6):1094-104.

\bibitem{venable2017lipid}
Venable RM, Ing{\'o}lfsson HI, Lerner MG, Perrin~Jr BS, Camley BA, Marrink SJ, et~al.
\newblock Lipid and Peptide Diffusion in Bilayers: The Saffman--Delbr\"uck Model and Periodic Boundary Conditions.
\newblock The Journal of Physical Chemistry B. 2017;121(15):3443-57.

\bibitem{shi2022hydrodynamics}
Shi W, Moradi M, Nazockdast E.
\newblock Hydrodynamics of a single filament moving in a spherical membrane.
\newblock Physical Review Fluids. 2022;7(8):084004.

\bibitem{huang2024interfacial}
Huang Y, Suja VC, Yang M, Malkovskiy AV, Tandon A, Colom A, et~al.
\newblock Interfacial stresses on droplet interface bilayers using two photon fluorescence lifetime imaging microscopy.
\newblock Journal of Colloid and Interface Science. 2024;653:1196-204.

\bibitem{evans1994hidden}
Evans E, Yeung A.
\newblock Hidden dynamics in rapid changes of bilayer shape.
\newblock Chemistry and Physics of Lipids. 1994;73(1-2):39-56.

\bibitem{edidin1976measurement}
Edidin M, Zagyansky Y, Lardner T.
\newblock Measurement of membrane protein lateral diffusion in single cells.
\newblock Science. 1976;191(4226):466-8.

\bibitem{saffman1975brownian}
Saffman P, Delbr{\"u}ck M.
\newblock Brownian motion in biological membranes.
\newblock Proceedings of the National Academy of Sciences. 1975;72(8):3111-3.

\bibitem{evans1988translational}
Evans E, Sackmann E.
\newblock Translational and rotational drag coefficients for a disk moving in a liquid membrane associated with a rigid substrate.
\newblock Journal of Fluid Mechanics. 1988;194:553-61.

\bibitem{haidekker2007molecular}
Haidekker MA, Theodorakis EA.
\newblock Molecular rotors—fluorescent biosensors for viscosity and flow.
\newblock Organic \& Biomolecular Chemistry. 2007;5(11):1669-78.

\bibitem{nipper2008characterization}
Nipper ME, Majd S, Mayer M, Lee JCM, Theodorakis EA, Haidekker MA.
\newblock Characterization of changes in the viscosity of lipid membranes with the molecular rotor FCVJ.
\newblock Biochimica et Biophysica Acta (BBA)-Biomembranes. 2008;1778(4):1148-53.

\bibitem{levitt2009membrane}
Levitt JA, Kuimova MK, Yahioglu G, Chung PH, Suhling K, Phillips D.
\newblock Membrane-bound molecular rotors measure viscosity in live cells via fluorescence lifetime imaging.
\newblock The Journal of Physical Chemistry C. 2009;113(27):11634-42.

\bibitem{wu2013molecular}
Wu Y, {\v{S}}tefl M, Olzy{\'n}ska A, Hof M, Yahioglu G, Yip P, et~al.
\newblock Molecular rheometry: direct determination of viscosity in L o and L d lipid phases via fluorescence lifetime imaging.
\newblock Physical Chemistry Chemical Physics. 2013;15(36):14986-93.

\bibitem{lopez2014molecular}
L{\'o}pez-Duarte I, Vu TT, Izquierdo MA, Bull JA, Kuimova MK.
\newblock A molecular rotor for measuring viscosity in plasma membranes of live cells.
\newblock Chemical Communications. 2014;50(40):5282-4.

\bibitem{dent2015imaging}
Dent MR, L{\'o}pez-Duarte I, Dickson CJ, Geoghegan ND, Cooper JM, Gould IR, et~al.
\newblock Imaging phase separation in model lipid membranes through the use of BODIPY based molecular rotors.
\newblock Physical Chemistry Chemical Physics. 2015;17(28):18393-402.

\bibitem{forster1971viskositatsabhangigkeit}
F{\"o}rster T, Hoffmann G.
\newblock Die viskosit{\"a}tsabh{\"a}ngigkeit der fluoreszenzquantenausbeuten einiger farbstoffsysteme.
\newblock Zeitschrift f{\"u}r Physikalische Chemie. 1971;75(1\_2):63-76.

\bibitem{haidekker2010environment}
Haidekker MA, Theodorakis EA.
\newblock Environment-sensitive behavior of fluorescent molecular rotors.
\newblock Journal of Biological Engineering. 2010;4(1):11.

\bibitem{kuimova2012mapping}
Kuimova MK.
\newblock Mapping viscosity in cells using molecular rotors.
\newblock Physical Chemistry Chemical Physics. 2012;14(37):12671-86.

\bibitem{singh2023molecular}
Singh G, George G, Raja SO, Kandaswamy P, Kumar M, Thutupalli S, et~al.
\newblock A molecular rotor FLIM probe reveals dynamic coupling between mitochondrial inner membrane fluidity and cellular respiration.
\newblock Proceedings of the National Academy of Sciences. 2023;120(24):e2213241120.

\bibitem{adrien2022best}
Adrien V, Rayan G, Astafyeva K, Broutin I, Picard M, Fuchs P, et~al.
\newblock How to best estimate the viscosity of lipid bilayers.
\newblock Biophysical Chemistry. 2022;281:106732.

\bibitem{anthony2022systematic}
Anthony AA, Sahin O, Yapici MK, Rogers D, Honerkamp-Smith AR.
\newblock Systematic measurements of interleaflet friction in supported bilayers.
\newblock Biophysical Journal. 2022;121(15):2981-93.

\bibitem{zgorski2019surface}
Zgorski A, Pastor RW, Lyman E.
\newblock Surface shear viscosity and interleaflet friction from nonequilibrium simulations of lipid bilayers.
\newblock Journal of Chemical Theory and Computation. 2019;15(11):6471-81.

\bibitem{camley2013diffusion}
Camley BA, Brown FL.
\newblock Diffusion of complex objects embedded in free and supported lipid bilayer membranes: role of shape anisotropy and leaflet structure.
\newblock Soft Matter. 2013;9(19):4767-79.

\bibitem{sackmann1996supported}
Sackmann E.
\newblock Supported membranes: scientific and practical applications.
\newblock Science. 1996;271(5245):43-8.

\bibitem{stone1998hydrodynamics}
Stone HA, Ajdari A.
\newblock Hydrodynamics of particles embedded in a flat surfactant layer overlying a subphase of finite depth.
\newblock Journal of Fluid Mechanics. 1998;369:151-73.

\bibitem{oppenheimer2010correlated}
Oppenheimer N, Diamant H.
\newblock Correlated dynamics of inclusions in a supported membrane.
\newblock Physical Review E. 2010;82(4):041912.

\bibitem{sluch2002friction}
Sluch MI, Somoza MM, Berg MA.
\newblock Friction on small objects and the breakdown of hydrodynamics in solution: rotation of anthracene in poly (isobutylene) from the small-molecule to polymer limits.
\newblock The Journal of Physical Chemistry B. 2002;106(29):7385-97.

\bibitem{evans1976membrane}
Evans E, Hochmuth R.
\newblock Membrane viscoelasticity.
\newblock Biophysical Journal. 1976;16(1):1-11.

\end{thebibliography}
\end{document}